\documentclass[3p]{elsarticle} 

\usepackage{lineno,hyperref}
\modulolinenumbers[5]
\usepackage{subcaption}
\usepackage{multirow}
\usepackage{colortbl}
\usepackage[table,xcdraw]{xcolor}
\usepackage{bm}
\usepackage{enumitem}
\usepackage{amsmath}
\usepackage{listings}

\makeatletter
\def\ps@pprintTitle{%
   \let\@oddhead\@empty
   \let\@evenhead\@empty
   \def\@oddfoot{\reset@font\hfil\thepage\hfil}
   \let\@evenfoot\@oddfoot
}
\makeatother

\begin{document}

\begin{frontmatter}

\title{Extracting the signed backbone of intrinsically dense weighted networks}

\author{Furkan~Gürsoy\corref{cor1}}
\ead{furkan.gursoy@boun.edu.tr}

\author{Bertan~Badur}
\ead{bertan.badur@boun.edu.tr}

\cortext[cor1]{Corresponding author}

\address{Dept. of Management Information Systems, Boğaziçi University, Istanbul 34342,Turkey}

\begin{abstract}
Networks provide useful tools for analyzing diverse complex systems from natural, social, and technological domains. Growing size and variety of data such as more nodes and links and associated weights, directions, and signs can provide accessory information. Link and weight abundance, on the other hand, results in denser networks with noisy, insignificant, or otherwise redundant data. Moreover, typical network analysis and visualization techniques presuppose sparsity and are not appropriate or scalable for dense and weighted networks. As a remedy, network backbone extraction methods aim to retain only the important links while preserving the useful and elucidative structure of the original networks for further analyses. Here, we provide the first methods for extracting signed network backbones from intrinsically dense unsigned unipartite weighted networks. Utilizing a null model based on statistical techniques, the proposed \textit{significance filter} and \textit{vigor filter} allow inferring edge signs. Empirical analysis on migration, voting, temporal interaction, and species similarity networks reveals that the proposed filters extract meaningful and sparse signed backbones while preserving the multiscale nature of the network. The resulting backbones exhibit characteristics typically associated with signed networks such as reciprocity, structural balance, and community structure. The developed tool is provided as a free, open-source software package.
\end{abstract}


\end{frontmatter}

\section{Introduction}

Networks are increasingly useful in modeling and studying many problems in seemingly unrelated domains from social sciences \cite{socialsciences}, natural sciences and engineering \cite{naturalsciences}, and arts and humanities \cite{humanities}.
A very simple network consists of nodes (vertices, actors) and links (edges, ties) connecting them. 
The network science tools utilize the information from the interdependence between entities given by the network structure. 
The data that can be modeled in the network form is not limited only to the node and edge structure. More complex networks have attributes associated with edges and nodes, especially with the advent of advanced data collection methods.

In weighted networks, edges have numeric weights indicating their intensity (e.g., coupling strength, amount, similarity, etc.). In directed networks, edges have distinguishable source and target nodes, indicating the direction of the relation (e.g., flow, following, etc.). For instance, the air traffic can be modeled as a very simple network such that a node represents an airfield and an edge between two nodes indicates a flight between them. In this case, the number of connections an airfield has is given by the degree of the corresponding node. The same system can also be modeled as a directed weighted network where an edge carries the take-off and landing airfield data as its direction and passenger capacity data as its weight. In this way, the outgoing and incoming capacity of an airfield is given by the out- and in-strength of the node, that is the total weight of edges leaving and entering the node, respectively. This very simplified example demonstrates the view that as networks get more complex, richer analyses can be made.

Another type of networks, which is explored relatively less in the literature, are signed networks where edges have negative or positive signs respectively indicating antagonism (e.g., dislike, distrust, foes, dissimilarity, voting against, inhibition, etc.) or rapport (e.g., like, trust, friendship, similarity, voting for, activation, etc.). Negative edges are not simply negation of positive edges \cite{negation}, they show distinct behavior in networks \cite{Szell13636}, and their inclusion enrich the typical tasks on networks 
such as link prediction \cite{kunegis, GUPTA2020113321}, 
recommender systems \cite{jiliangrecom, ZHANG2019317}, 
node classification \cite{mercadonodeclass, tangnodeclass},
node centrality and ranking \cite{gangal2016trust, Wan2019}, 
representation learning \cite{kimRLSIDE, derrRLGCN}, 
information diffusion and influence maximization \cite{HOSSEINIPOZVEH2019476, JU20201571}, 
finding cliques \cite{liCliq},
community detection, graph partitioning and blockmodels \cite{esmacommun, MAcomm, liublock}, 
and polarization \cite{bonchipolarization, xiaopolarization}. In addition to the typical and prevalent applications in social media \cite{tangSMsurvey}, signed networks also find application areas in
politics \cite{arinikparliament}, 
international relations \cite{DoreianINTREL},
finance \cite{hararyPortfolio}, 
biology \cite{iaconoSysBio, ouyangPPI, huDrugSide}, 
and ecology \cite{huDrugSide}.

The density of a network is usually defined by the ratio of the number of observed links to the number of possible links. Formally, sparse networks are those where density asymptotically goes to zero in the limit of large number of nodes. However, in most empirical networks, it is impossible to evaluate this \cite{newman2018networks}. As a result, in general, an empirical network is called sparse if it has a low density. The majority of real-world networks are sparse \cite{newman2003structure}. For instance, the number of stable social relations of a human is limited by cognitive constraints \cite{dunbarvespignani}, stations are usually only connected to nearest stations in power grid networks, a webpage points out only so many others in web networks, and so on. On the other hand, some networks might be very dense or even almost completely connected by their own nature. Most countries trade with most other countries, humans move/migrate from many locations to many other locations, many species have many predators and preys \cite{DunneFoodweb}, most people interact with most others in certain social settings, and so on.

The earlier view that additional data allowing richer analyses is restricted by the amount and structure of the data itself, particularly in the case where the additional data is provided by more links between the nodes. In addition to the fact that sparsity is desirable due to the computational complexity of many network algorithms, most of the typical network analysis and visualization methods assume the networks to be sufficiently sparse \cite{newman2018networks}. Apart from computational and methodological concerns, dense networks might have noisy, uninformative, insignificant, or otherwise redundant links. This further worsens the application and interpretation capacity in many network tasks, for instance, those relating to node centrality, cliques and communities, and diffusion. Therefore, the relevant and significant information should be extracted from such dense networks such that the original rich data is reduced into a network that is sparse and simpler but maintains adequate structural information for efficient and effective analyses.

There is a body of literature for extracting backbones of usually weighted and relatively dense networks, which we review in the next section. Such information filtering task is usually referred to as backbone extraction or network sparsification and aims to remove statistically insignificant or otherwise redundant links while maintaining the informative structural properties for further analysis. To the best of our knowledge, this is the first study in the literature that provides methods for extracting signed network backbones from weighted, dense, and originally unipartite networks. Unlike the earlier studies, our method relies on and requires the input network to be intrinsically dense; which in turn enables inferring the link signs.

We use the term \textit{intrinsically dense} for characterizing networks where all nodes, in a sense, are aware of all other nodes and can interact with them without obvious natural limits such as those mentioned earlier for sparse networks. This definition follows that an edge does not necessarily represent the existence of a positive relationship but might be an artifact of the randomness or even a negative relationship depending on its weight. Such definition of intrinsically dense networks provides a distinction not only from sparse networks but also from certain dense networks. For instance, a human interaction network in a workshop with two parallel sessions with a single mutual short break is not intrinsically dense because the participants in different sessions are not given sufficient time to meet with others, thus, the absence of a link cannot reasonably indicate if two participants from the different sessions are avoiding each other. As a demonstrating example of intrinsically dense networks, consider a network where link weights denote the similarity of respective nodes. Lower similarity does not only mean a lack of similarity but often indicates dissimilarity, a negative underlying link. In the same vein, voting for a set of candidates indicates support for those candidates while also suggesting opposition for the other candidates especially those who are otherwise popular. Finally, we should highlight that whether a network is intrinsically dense or not is a matter of extent and not a strictly binary decision.

We suggest that positive and negative links are those with weights significantly and substantially deviating from the random expectation under a suitable null model. We develop an appropriate null model based on hypergeometric distribution and iterative proportional fitting procedure and offer \textit{significance filter} and \textit{vigor filter} for extracting signed backbones of intrinsically dense networks. The proposed methodology is capable of handling directed or undirected networks as input and producing weighted or unweighted signed backbones.

Our contribution can be summarized as follows.
\vspace{-5pt}
\begin{itemize}[itemsep=0pt]
    \item An extensive literature review on network backbone extraction methods is provided (Section \ref{sec:relwork}).
    \item The first methods to extract signed backbones of intrinsically dense, weighted, originally unipartite networks are proposed (Section \ref{sec:methods}).
    \item The methodology is empirically evaluated on real-world networks of different characteristics and its feasibility and usefulness is shown (Section \ref{sec:empiric}).
    \item The proposed signed network backbone extraction tool is provided as an open-source software package (Appendix A).
\end{itemize}

\section{Backbone extraction methods}\label{sec:relwork}

The simplest yet a popular approach is to apply a global threshold where only those edges with weights satisfying the predetermined threshold are retained. There are two major problems with this approach. First, the choice of the threshold is rather arbitrary and non-impartial unless the physical meaning of the weights in the domain of the particular network allows an explanation. The second and more important problem is the case of multiscale networks where the weights are distributed over a broad range of scales. Such characteristic is definitely not an exception but an observed phenomenon in many real-world networks \cite{Barrat2004}. For instance, in many networks where rich club effects \cite{mondragon2004} are prevalent, nodes with higher degrees/strengths tend to form links with nodes of the same or higher degree/strength which results in a hierarchical and multiscale network structure. Rich club effects and such structures are present in many weighted networks including the global airline traffic network \cite{alstott2014unifying}, worldwide maritime transportation network \cite{HU20092061}, world trade network \cite{Bhattacharya_2008, schiavo2010}, international credit-debt networks \cite{CHINAZZI20131692}, population flow network during a national holiday \cite{WEI201877}, patient referral networks \cite{TANG2020492}, and  communication networks in the brains of humans \cite{alstott2014unifying} and rats \cite{ratconnectome}. Therefore, weights with small magnitudes are not necessarily noise and weights with large magnitudes are not necessarily important but might be the result of such multiscale weight characteristics. In multiscale networks, the application of a global threshold tends to eliminate the edges of low-strength nodes regardless of the local significance of those edges. As a result, the structure of high-strength nodes can be preserved but local regions characterized by relatively low-strength nodes are underestimated, might become disconnected, and might even be completely eliminated. In summary, increasing the value of a global threshold loses local structures at the bottom of the hierarchy whereas decreasing it results in retained noise in the upper levels. Consequently, global thresholding would only work under the often-unrealistic assumption that edge weights are independent and identically distributed random variables.

As a remedy, the global threshold method can be improved to work on weights represented not in universal units but as fractions of the node degrees/strengths. Yet, an edge is associated with two nodes and it again becomes relatively arbitrary to choose the normalization factor. Such normalization might not be robust to nodes with very low and high degrees as well since it potentially overestimates the edges of low-degree nodes and underestimates the edges of high-degree nodes. Overall, it could serve as a practical remedy for certain problems but the need for more statistically sound methods is clear. Another approach is to extract the spanning tree with maximal weight (MST) by appropriately transforming the weights and finding the minimum spanning tree. This method does not require any extrinsic input and ensures that the extracted backbone is connected. However, the backbone will be acyclic and local clustering and the community structure cannot be preserved. Using a similar spanning tree approach, Tumminello et al. \cite{Tumminello10421} propose a technique for controlling the genus of the backbone which, in turn, enables preservation of some local structure.

Grady et al. \cite{Grady2012} extract the \textit{high-salience skeleton} of the network which consists of only the salient edges. For each node, they construct the shortest path tree by merging all shortest paths from the node to all other nodes. Then, the saliency of an edge is defined as the fraction of all shortest-path trees where the edge is a member. They show that edge saliency shows a bimodal distribution near the boundaries. Only the edges with saliency near $1$ are retained; effectively eliminating the need for choosing an arbitrary threshold. By design, it ensures the connectivity of the extracted backbone. Edge saliency is different from edge betweenness where the former tends to award the edges in the periphery (e.g., low-degree nodes) whereas the latter tends to award the edges in the core (e.g., high-degree nodes).

The methods described so far do not assume any underlying null model and do not compare observed weights to expected weights for statistical evaluation. \textit{Disparity filter} \cite{Serrano2009backbone} assumes that the normalized weights of the links of a node follow a uniform distribution. Comparing the observed weights to this null model at a desired significance level, the network backbone including only the statistically significant links can be obtained. As mentioned earlier, a weight can be normalized and evaluated separately for the two nodes it connects. Thus, a link can be significant from the viewpoint of one node and not the other. This problem is tackled by retaining a link if it satisfies the significance condition for at least one of the two nodes. Adopting a loosely similar weight normalization scheme, \textit{bistochastic filter} \cite{SlaterE66} first scales the weight matrix such that its marginals (i.e., the row and column totals, in- and out-strength sequence) are equal to $1$ resulting in a doubly stochastic matrix. Then, the links with the highest weights are retained until the network is strongly connected; or until any other stopping criteria which can result in sparser backbones with disconnected components or even denser backbones.

In the same line of research, \textit{LANS} \cite{foti2011nonparam} does not assume an underlying distribution for the weights of a node and employs the empirical cumulative density function instead. It is stated that it is more robust to the highly heterogeneous local weight distributions than \textit{disparity filter} and \textit{bistochastic} filter are.  Rather than adopting a local approach in developing null models, \textit{GloSS} filter \cite{Radicchi2011} utilizes a global null model. The model considers both the strengths and degrees and aims to preserve the global weight distribution and the topology. \textit{GLANB} \cite{zhang2014extracting} evaluates the statistical significance of link saliency to retain satisfactory links; loosely combining \textit{disparity filter} and \textit{high-salience skeleton}.

Dianati et al. \cite{dianati2016hairball}  introduce two interrelated filters. Treating an integer-weighted network as a multiedge network, they assume that each unit edge randomly chooses two nodes respecting the degree sequences which results in a binomial distribution for weights. The null model is indicated to blend the approaches of  \textit{disparity filter} and \textit{GloSS}. \textit{Marginal likelihood filter} operates by evaluating each edge separately against a chosen significance level whereas \textit{global likelihood filter} incorporates exponential random graph model with a Monte Carlo simulation scheme to consider all links at once. \textit{ECM filter} \cite{gemmetto2017irreducible}  enhances the null model of Dianati et al. \cite{dianati2016hairball} with the purpose of retaining the relation between strengths and degrees. \textit{Polya filter} \cite{Marcaccioli2019} assumes an underlying null model with a self-reinforcing mechanism in which the generation of link weight increases the probability of further weight generations for that link. 

Tumminello et al. \cite{tummihypergeom} introduce \textit{hypergeometric filtering} originally for bipartite weighted networks. It is employed for unipartite weighted networks \cite{Riccaboni2013, SGRIGNOLI2015245} as well. With fixed node in- and out-strengths, assuming integer weights and treating weighted links as multiple links, a unit-weight link from a node chooses its other end randomly among all possible nodes. Then, the edge weight generation process can be described by a hypergeometric distribution which becomes the null model for the filter. Comparing observed values with the null model, statistically significant links can be extracted at desired levels of confidence. \textit{Noise-Corrected Bayesian filter} \cite{coscia2017backboning} assumes edge weight generation follows a binomial distribution where a null model based on the hypergeometric distribution is used in determining Bayesian priors. Lift value for each link (i.e., the ratio of the observed value to the expected value) is calculated, transformed to the $[-1, 1]$ range, and the associated variance is estimated with a Bayesian inference schema. Using the appropriate posterior variances, the links which satisfy a desired significance level are retained.

The studies described so far except for \cite{tummihypergeom} are primarily designed for and concerned with unipartite networks. When backbones are extracted from the one-mode projections of originally bipartite networks, a major concern is the loss of relevant and important information. A binary bipartite network where links represent the relations (usually events) between two different types of nodes (e.g., agents and artifacts) can be projected into a one-mode network where link weights between agents denote the number of shared artifacts. However, such projection loses some information such as the degree distribution of the artifacts and which artifacts are common between any given two agents. Neal \cite{neal2014} proposes stochastic degree sequence model that generates a reasonably large set of random bipartite networks conditioned on both agent and artifact degrees. Then, expected weight distributions from the projections of generated networks are employed for testing the significance and sign of links in the original projection. Liebig and Rao \cite{liebig2016} show that edge weights in such projections follow a Poisson binomial distribution and identifies the significant links without generating random bipartite networks. Domagalski et al. \cite{domagalski2021} provide a software package for extracting the binary or signed backbones of projections of bipartite networks using hypergeometric model, stochastic degree sequence model, or fixed degree sequence model. Overall, these methods are better suited when the backbone is to be extracted from one-mode projections of bipartite binary networks and the information on the original bipartite network is available. Comprehensive reviews on backbones of bipartite projections can be found in \cite{neal2014, domagalski2021}.

There are more specific approaches with different purposes, e.g., to retain the most consistent links in brain networks based on diffusion images \cite{ROBERTS2017118}, to preserve functional backbones based on network motifs \cite{CAO2019motif}), and to extract underlying networks from correlation matrices \cite{scola}. Generally, the state-of-the-art concentrates on developing appropriate null models to evaluate link weights against and retain only those satisfying a desired level of statistical significance.

\section{Methods}\label{sec:methods}

\subsection{Formal Problem Definition}

We denote an intrinsically dense, undirected or directed, and non-negative weighted network without self-loops by $G := (V, W)$. $V$ is the set of nodes with $i, j \in V$ as its general elements and its cardinality is $n = |V|$.  $W$ is the corresponding weight matrix. When $G$ is undirected $W_{ij}$ denotes the weight of the link between $i$ and $j$ and $W_{ij} = W_{ji}$. When $G$ is directed, $W_{ij}$ denotes the weight of the link from $i$ to $j$. Self-loops are not allowed hence $W_{ii} = 0  \; \forall i$. Total outflow of $i$ is denoted with $W_{i.} = \sum_{j} W_{ij}$. Total inflow to $j$ is denoted with $W_{.j} = \sum_{i} W_{ij}$. The sum of all weights is denoted with $W_{..} = \sum_{i} \sum_{j} W_{ij}$. We employ the same definitions of $W_{i.}$, $W_{.j}$, and $W_{..}$ for directed and undirected networks, i.e., treating an undirected link as two reciprocal directed links with equal weights.

Given $G$, the aim is to extract a meaningful signed network $\hat{G} := (\hat{V}, \hat{A}, \hat{W})$ where $\hat{A}$ is a sparse adjacency matrix in general such that $\hat{A}_{ij} \in \{-1, 0, 1\}$ where $-1$ denotes a negative link, $1$ denotes a positive link, and $0$ indicates the absence of a link. $\hat{W}$ is the corresponding optional weight matrix. If $\hat{A}_{ij} = 0$, then $\hat{W}_{ij} = 0$; otherwise $\hat{W}_{ij}$ has the same sign as $\hat{A}_{ij}$.

\subsection{Proposed Solution}

\textbf{The null model.} 
To extract the signed edges, it is necessary to have a null model which empirical edge weights are compared against to distinguish edge weights that are due to chance and edge weights that significantly deviate from the expected values in either positive or negative direction. Our null model ensures that node strengths are fixed and edge weight distributions are characterized by a sampling without replacement procedure.

We assume that in- and out- strengths are fixed, thus, each node has a specified number of stubs (i.e., half edges). A unit-link is a connection of two stubs from different nodes. When a stub makes a connection, it chooses the other stub randomly from all available stubs. Hence, for a stub, the probability of connecting to a specific node is proportional number of its available stubs. The weight of the link between two nodes is equal to the number of unit-links between them. In this way, the process is reduced to sampling without replacement problem, of which the urn problem is a famous example. We describe the process with a simple analogy. For each node $i$, there are $W_{..} - W_{.i}$ marbles in the urn (not allowing self-loops), $W_{i.}$ marbles are chosen from the urn without replacement, and there are $W_{.j}$ marbles in the urn for each $j$. Such process is well-characterized by hypergeometric distribution and allows us to calculate statistical quantities for links between all $i$ and $j$. 

The mean of the hypergeometric distribution associated with the link from $i$ to $j$, that is $P_{ij}$, is given by Eq. \ref{eq:hgmean}. However, this formulation does not preserve the in-strength and out-strength sequences in the system, thus, is inadequate for our purposes. Given that $N$ is the weight matrix under the null model with $N_{ij}$ denoting the expected weight from $i$ to $j$, the following should be satisfied: $N_{i.} = W_{i.}$, $N_{.j} = W_{.j}$, and $N_{..} = W_{..}$. For this purpose, we utilize the iterative proportional fitting procedure (IPFP) described in \cite{Bacharach, mazzarisi2017methods}. Given out- and in-strength ($W_{i.}$ and $W_{.j}$) sequences as the marginals and $P$ as the prior matrix with its diagonal elements set to $0$, the objective of IPFB is to estimate $N$ that minimizes the relative entropy (known as Kullback–Leibler divergence \cite{kl_divergence}) as formalized in Eq. \ref{eq:entropy}. IPFP estimates the values for non-zero elements of the matrix through a series of row-scaling and column-scaling operations until achieving a desired precision and is shown to converge to an optimal solution \cite{BREGMAN1967191}. Then, each element of $N$ serves as the expected value of the respective link under our null model.

\begin{equation}
    P_{ij} = \frac{W_{i.} W_{.j}}{W_{..} - W_{.i}}
\label{eq:hgmean}
\end{equation}

\begin{equation}
\begin{aligned}
\min _{N_{i j}} \sum_{i, j \neq i} N_{i j} \log \frac{N_{i j}}{P_{i j}} \\
\text{subject to } N_{i.} = W_{i.}\text{, } N_{.j} = W_{.j}\text{, } N_{i j} \geq 0\text{, } N_{i i} = 0\text{; } \forall i, j
\end{aligned}
\label{eq:entropy}
\end{equation}

In evaluating the statistical significance of an edge weight, only considering the difference of observed and expected values without referencing a dispersion measure is not viable. For each edge, dispersion of its expected weight must be known to produce confidence intervals. Accordingly, we estimate the variance of $N_{ij}$ based on the associated hypergeometric distribution and obtain its well-known standard deviation as given by Eq. \ref{eq:hgstd}.

\begin{equation}
\sigma_{ij} = \sqrt{\left( W_{i.} \; \frac{W_{.j}}{W_{..} - W_{.i}} \; \frac{(W_{..} - W_{.i}-W_{.j})}{W_{..} - W_{.i}} \; \frac{W_{..} - W_{.i}-W_{i.}}{W_{..} - W_{.i}-1} \right)}
\label{eq:hgstd}
\end{equation}

Overall, our null model can be viewed as extending the hypergeometric approach employed in the literature to intrinsically dense unipartite networks and to ensure that node strengths are exactly preserved with the inclusion of entropy-based method IPFP. Moreover, distinctly from the existing hypergeometric filtering literature, we do not necessarily assume weights to be integer counts of events hence we do not employ the traditional hypothesis testing procedure associated with the hypergeometric distribution.

\textbf{Significance filter.} 
The first filter we propose functions to eliminate links whose weights do not significantly deviate from their expected values. \textit{Significance filter} eliminates those links satisfying the condition $\sigma_{ij} \: \alpha^- \: < W_{ij} - N_{ij} < \: \sigma_{ij} \: \alpha^+$ where $\alpha^-$ and  $\alpha^+$ respectively take non-positive and non-negative values and are user-defined hyperparameters specifying the desired significance thresholds for negative and positive signed edges.

It should be highlighted that the hypergeometric distribution is a distribution for discrete events. In general, we can slightly abuse it by allowing continuous weights or rounding weights to the nearest integers without much impact on the resulting backbone. However, a larger issue that is not adequately discoursed in the hypergeometric filtering literature is the impact of the magnitude of weights on the statistical significance evaluation. As the magnitude of weights increases, the confidence intervals relative to the weights become narrower, as a consequence of the assumption that the weights are the counts of discrete events. A demonstrating example is provided in Appendix B. Therefore, for instance, the same monetary network represented in dollars or cents would result in different network backbones when they are evaluated at the same statistical significance level. This issue is less pressing when the link weights essentially correspond to discrete events such as human migration networks or voting networks. Taking these into account, it can be concluded that the values chosen for $\alpha$ might not accurately translate into traditional \textit{p-values} in many networks due to their weights' non-discrete and arbitrary-unit nature. Hence, the choice of $\alpha$ should not necessarily be limited to traditionally employed values especially when the weights do not correspond to discrete events.

Alternatively, since the plausible range of $\alpha$ values may vary in networks with different magnitudes of weights, one may filter the backbone based on edge ranking. Such rank-based filtering was employed in the above-discussed literature as well \cite{SlaterE66, dianati2016hairball, Marcaccioli2019}. Instead of providing significance threshold values, a backbone can be extracted in a way that it has a desired level of sparsity. One may choose to retain top $x\%$ of edges based on their absolute significance levels (i.e., $|W_{ij} - N_{ij}|/\sigma_{ij}$) or bottom $y\%$ and top $z\%$ of edges based on (non-absolute) significance levels. Ultimately, any rank-based filtering is equivalent to some $(\alpha^-, \alpha^+)$ configuration but may be more intuitive.

\textbf{Vigor filter.} 
The significance filter might retain statistically significant but otherwise very weak links. Yet, an edge can be evaluated also based on whether it is sufficiently strong in terms of intensity. This is useful for multiple reasons: (i) the employed \textit{significance filter} might be too permissive in certain circumstances, (ii) the signed network backbone is needed to reflect only binary links where opinions are rather strong, or (iii) higher sparsity is desired. We define vigor of an edge from $i$ to $j$ as $\beta_{ij}$ (pronounced as \textit{víta} in Modern Greek) in Eq. \ref{eq:vigor}. Vigor values may be considered as edge weights normalized based on their expected values under our null model. It takes values in the range $[-1, 1]$, its magnitude indicates the intensity of the link and $\beta_{ij} = 0$ when  $W_{ij} = N_{ij}$. \textit{Vigor filter} eliminates those links satisfying the condition $\beta^- < \beta_{ij} < \beta^+$ where $\beta^+$ and $\beta^-$ respectively take non-positive and non-negative values and are user-defined hyperparameters  specifying the desired vigor thresholds for positive and negative signed edges.

\begin{equation}
    \beta_{ij} = \frac{W_{ij}/N_{ij} \; -1}{W_{ij}/N_{ij} \; +1}
\label{eq:vigor}
\end{equation}

\textbf{Signed backbone extraction.}
The signed backbone of a network can be extracted using a combination of \textit{significance filter} and \textit{vigor filter}. The former serves as the primary filter and should always be employed since $\beta$ values do not reflect the statistical significance and might be unreliable in the small-magnitude weight regime characterized by small $N_{ij}$, relatively large $\sigma_{ij}$, and small $W_{ij}$. \textit{Significance filter} would function to eliminate such links, therefore, the retained links would have reliable vigor. Vigor values of the remaining edges can be utilized as the signed edge weights in the extracted backbone. In the case of weighted backbone, the use of \textit{vigor filter} is rather optional since the vigor information is directly carried onto the network. 

Alternatively, the weights can be ignored to produce a network with binary opinions. In this case, one may consider utilizing also the \textit{vigor filter} to retain only the edges that exhibit appropriate levels of intensity. In the high-magnitude regime characterized by large $N_{ij}$, relatively small $\sigma_{ij}$, and large $W_{ij}$; statistical significance may be easier to obtain. Accordingly, we may confidently identify a negative or positive relationship without necessarily uncovering whether the relation is intense. This is intuitive since, for instance, a  slightly positive or negative relationship between two actors can be identified more confidently when we have more observations on them while higher confidence does not necessarily mean a higher intensity of rapport or antagonism. Since $\beta_{ij}$ values have interpretable physical meanings as they are comparable weights that were normalized based on null expectations (e.g., $\beta = 0.33$ indicates that the observed value is double the expected value), users may choose appropriate values of $\beta$ as they see fit for their problem.

Overall, one should always employ \textit{significance filter} to retain those links where the sentiment of a link can be confidently identified. Additionally, one may utilize \textit{vigor filter} to retain those links where the sentiment is rather strong. In setting appropriate values for the hyperparameters, one should consider (i) the physical meaning of hyperparameters in the context of the specific problem and (ii) the desired level of sparsity for the extracted backbone. In addition, one may look at the distribution of $\alpha_{ij}$ and $\beta_{ij}$ values for all edges before deciding on appropriate threshold values.

\textbf{Undirected networks.} The described method is defined well on the directed networks. The generalization to the undirected networks is ensured in the following way. An undirected link is replaced with two reciprocal directed links of the same weight; effectively transforming the network into a directed network. The null model and the filters produce the same null expectation and nearly the same variance\footnote{For sufficiently large networks, $W_{..} - W_{.j} \approx W_{..} - W_{.i}$ in In Eq. \ref{eq:hgstd}} for such reciprocal links.  Finally, the directed backbone can be transformed back into an undirected backbone by treating reciprocal directed links as undirected links and removing the redundancy by keeping the link with the highest absolute vigor.

\section{Empirical Analysis}\label{sec:empiric}

In this section, relevance and diverse characteristics of the employed network datasets are described, and proposed filters are empirically analyzed in terms of resulting backbone sizes, robustness to multiscale networks, and structures of the extracted signed network backbones.\footnote{The datasets and code for producing the analyses in this section are available at https://github.com/furkangursoy/signed\_backbones}

\subsection{Datasets.} 

The proposed method for signed network backbone extraction is experimented on four real-world networks from different domains and of varying statistical properties. For each network; the number of nodes $n$, number of nonzero links $m$, density, five-point summary of weights (minimum, first quantile, median, third quantile, and maximum values), and three-point summary (minimum, median, and maximum values) of node strengths (out- and in-strengths if networks are directed) are presented in Table \ref{tab:netstats}.

\begin{table}[h]
\small
\centering
\setlength{\tabcolsep}{0.5em}
\caption{Employed Networks}
\label{tab:netstats}
\setlength{\tabcolsep}{1pt}
\begin{tabular}{l|l|l|l|l|l|}
\cline{2-6}
 &
  \cellcolor[HTML]{EFEFEF}\bm{$n$} &
  \cellcolor[HTML]{EFEFEF}\bm{$m$} &
  \cellcolor[HTML]{EFEFEF}\textbf{density} &
  \cellcolor[HTML]{EFEFEF}\textbf{5-point summary of weights} &
  \cellcolor[HTML]{EFEFEF}\textbf{3-point summary of (out-, in-) strengths} \\ \hline
\multicolumn{1}{|l|}{\cellcolor[HTML]{EFEFEF}\textbf{Migration}}  & 51  & 2338 & 0.92 & (3, 356, 1079, 3236, 86k) & out:(21k, 101k, 691k), in:(26k, 109k, 587k) \\ \hline
\multicolumn{1}{|l|}{\cellcolor[HTML]{EFEFEF}\textbf{Eurovision}} & 26  & 260  & 0.4  & (1, 3, 6, 8, 12)          & out:(58, 58, 58), in:(0, 38, 167)            \\ \hline
\multicolumn{1}{|l|}{\cellcolor[HTML]{EFEFEF}\textbf{Contact}}    & 113 & 2196 & 0.17 & (1, 1, 2, 5, 1281)        & (2, 287, 1483)                              \\ \hline
\multicolumn{1}{|l|}{\cellcolor[HTML]{EFEFEF}\textbf{Species}}    & 62  & 1801 & 0.48 & (1, 26, 51, 76, 194)      & (954, 3249, 5448)                           \\ \hline
\end{tabular}
\end{table}

\textit{Migration} network \cite{uscensus} reports interstate migration flows (in terms of people) among the US states\footnote{The dataset contains District of Columbia in addition to the 50 states.} in 2018. Such internal migration data is explored for its correlation with economic productivity differences, geographical proximity, political preferences, and other cultural and historical factors \cite{CHAKRABARTI2017156, liu2019, Charyyev2019}. \textit{Eurovision} network \cite{eurovision} represents the votes between participant countries of the song contest in 2003\footnote{2003 was selected because it was the last year that all countries competed in a single round.}. Each country, via public votes\footnote{Ireland, Russia, and Bosnia and Herzegovina exercised jury voting instead of public voting.}, awards the set of points \{1,2,3,4,5,6,7,8,10,12\} to 10 other countries. Studies employing a network analysis perspective  \cite{FENN2006576, CHARRON2013484, garcia2013measuring, mantzaris2018collusion, Mantzaris2018preference, dangelo2019, svete2020just} show that voting behavior is not determined only by the music/performance quality but affected by political factors, geographical and cultural similarity, diasporas, and others. \textit{Contact} network data \cite{contact} is gathered via wearable sensors at an academic conference\footnote{The temporal face-to-face interaction dataset was collected at Hypertext conference in 2009 as part of \textit{SocioPatterns} research collaboration.} with 113 attendees over 2.5 days. The edge weights are generated by aggregating the number of 20-second intervals the respective two participants spent face-to-face over the course of the conference. The networks of such temporal ties are utilized for temporal backbone extraction, investigation of spreading processes, and other behavioral and structural analysis \cite{Kobayashi2019, Zhan2019, Kulisiewicz2018, Barrat2013}. \textit{Species} network \cite{species} is generated based on the cohabitation patterns of 62 marine species in South Florida. Edge weights represent the similarity of species based on the habitats they co-occupy during the same life stages\footnote{Given the species v. habitat-life stage bipartite matrix $B$, $W_{ij} := \sum_{k} B_{ik} B_{jk}$ where $i, j$ are the species and $k$ is the habitat-life stage pair, and $B_{ik}$ is the respective species-habitat-life stage score. More information on the original dataset can be found at https://atlanticfishhabitat.org/species-habitat-matrix/}. Information derived from such cohabitation networks is useful in various ecological studies \cite{kikkawa, camp2016cohabitation, Fernandes2020}. \textit{Species} network is generated from a bipartite network, thus, backbone extraction methods in the bipartite network projection literature are likely to be more suitable. However, it is also slightly different from the datasets in that literature since the original bipartite matrix here represents heterogeneous suitability scores instead of binary event-like relations. Moreover, none of the employed networks have natural self-loops; which makes them particularly suitable for our study.

Our proposed method considers the propensity of nodes for incoming and outgoing edge weights by explicitly including the in-strengths and out-strengths in the null model. For instance, high economic productivity for a state would result in more immigration into it. Similarly, low productivity would cause emigration from that state. The effect of such productivity is homogeneous in the sense that all other nodes perceive it in the same way. Such homogeneous effects are eliminated for the large part by our null model. However, for instance, the effect of political affiliation or geographic location is perceived differently by other nodes. Hence, the extracted backbone is expected to reflect non-economic factors in \textit{Migration} network. With a simple analogy, the quality of the songs and performances in the Eurovision song contest is equivalent to productivity. Hence, the elimination of such effect by the null model would result in an extracted signed network representing the non-quality factors in voting behavior. Similarly, the null model largely eliminates the effects of participants' tendency to engage in conversations (e.g., popularity, extroversion) and the extent of species' ability to live in many different habitats.

\subsection{Backbone Size}

In general, an extracted backbone should be sufficiently sparse such that only the important edges are preserved. At the same time, unlike edges, most nodes should be retained since they are often the subject of the analysis and are largely indispensable in understanding the global network structure. 

A few arguments for non-isolation of nodes in general cases are as follows. First, the very definition of networks includes relationships, thus, we expect nodes to have some underlying relations with other nodes particularly in intrinsically dense networks. Second, most real-world networks have high local clustering coefficient, which according to some studies \cite{granovetter1973strength, vanderLeijGoyal2011} are closely related to strong ties. Hence, it is generally safe to say that most nodes have at least some strong connections which should be retained by the backbone. Third, since we extract signed backbones, negative links also emerge which provide extra connectedness. Moreover, in the literature, MST and some other methods \cite{SlaterE66, Grady2012, Tumminello10421} ensure that backbones are fully connected while other studies \cite{coscia2017backboning, gemmetto2017irreducible, foti2011nonparam} present it as a generally desirable result.

Here, we extract the signed backbones of the four networks under various hyperparameter regimes and observe the size of the resulting backbone in terms of nodes and signed edges.

\begin{figure}[h]
    \centering
    \includegraphics[width=1\textwidth]{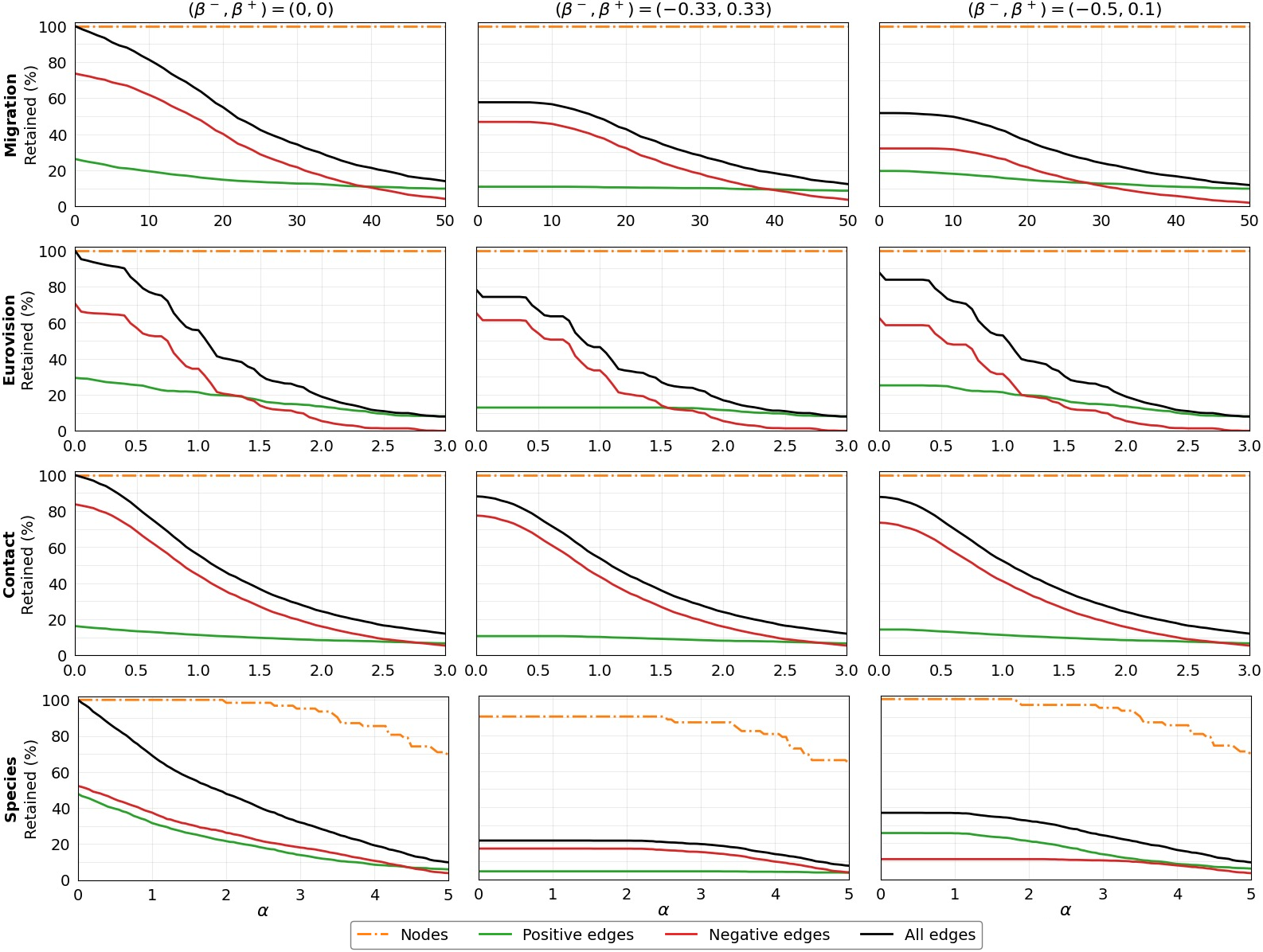}
    \caption{Effects of \textit{significance filter} on the backbone size.}
   
    \label{fig:retainalpha}
\end{figure}

\textbf{Effects of significance threshold.} Figure \ref{fig:retainalpha} visualizes the fraction of links and nodes retained for meaningful continuous ranges of the significance threshold under different vigor filters. 
The suitable range of $\alpha$ differs between networks mostly due to the varying overall magnitude of edge weights. The significance threshold is represented with $\alpha = \alpha^+ = - \alpha^-$ on \textit{x-axes}. Three different vigor threshold settings are represented in columns. For edges, percentage values on \textit{y-axes} are with respect to all possible edges, that is $n^2 - n$ edges for directed networks and half of it for undirected networks.

Except for \textit{Species} network, all nodes are retained in the backbones even under very strict settings where 80\% to 85\% of all possible edges are eliminated. 
When edges are inspected, apart from the \textit{Species} network where edges represent similarity, the negative edges are more frequent than the positive edges when $\alpha = 0, (\beta^-, \beta^+) = (0, 0)$. This is due to the case that edges with zero weights (i.e., nonexistent edges) or trivially small values are treated as negative links regardless of their statistical significance.
As the values of $\alpha$ increase, the number of negative links in the extracted backbone decreases faster than the positive links since many of the zero or trivially small weights are rather statistically insignificant.
On the other hand, utilizing vigor filter without the significance filter does not eliminate such zero-weighted links since those edges have vigor values of $-1$ (since $0$ divided by any null expectation is $0$, which maps to vigor of $-1$).
Hence, when $\alpha = 0$, the smaller number of negative edges in the second and third columns in comparison to the first column of Figure \ref{fig:retainalpha} is not due to the elimination of zero-weighted edges but the elimination of other weak edges. Therefore, \textit{significance filter} is always necessary when insignificant zero or trivially small edge weights exist.

Employing very large $\alpha$ values eliminates negative links almost completely. This is largely because the edges with relatively small expected values under the null model have relatively large variances. Increasing $\alpha$ too much, thus, tends to eliminate negative edges first as well as other edges in local regions characterized by edge weights of small magnitude. In the case of high-magnitude regions, an edge between two high-strength nodes is expected to have a large expectation under the null model. It is much harder for such edges to empirically observe much larger weights than the null expectation since the total weight in the network is fixed. This implies an upper boundary on vigor values for edges connecting to high-strength nodes. Therefore, increasing $\beta$ too much, especially on its positive range, tends to eliminate edges between central, high-strength nodes.

\begin{figure}[h]
    \centering
    \includegraphics[width=1\textwidth]{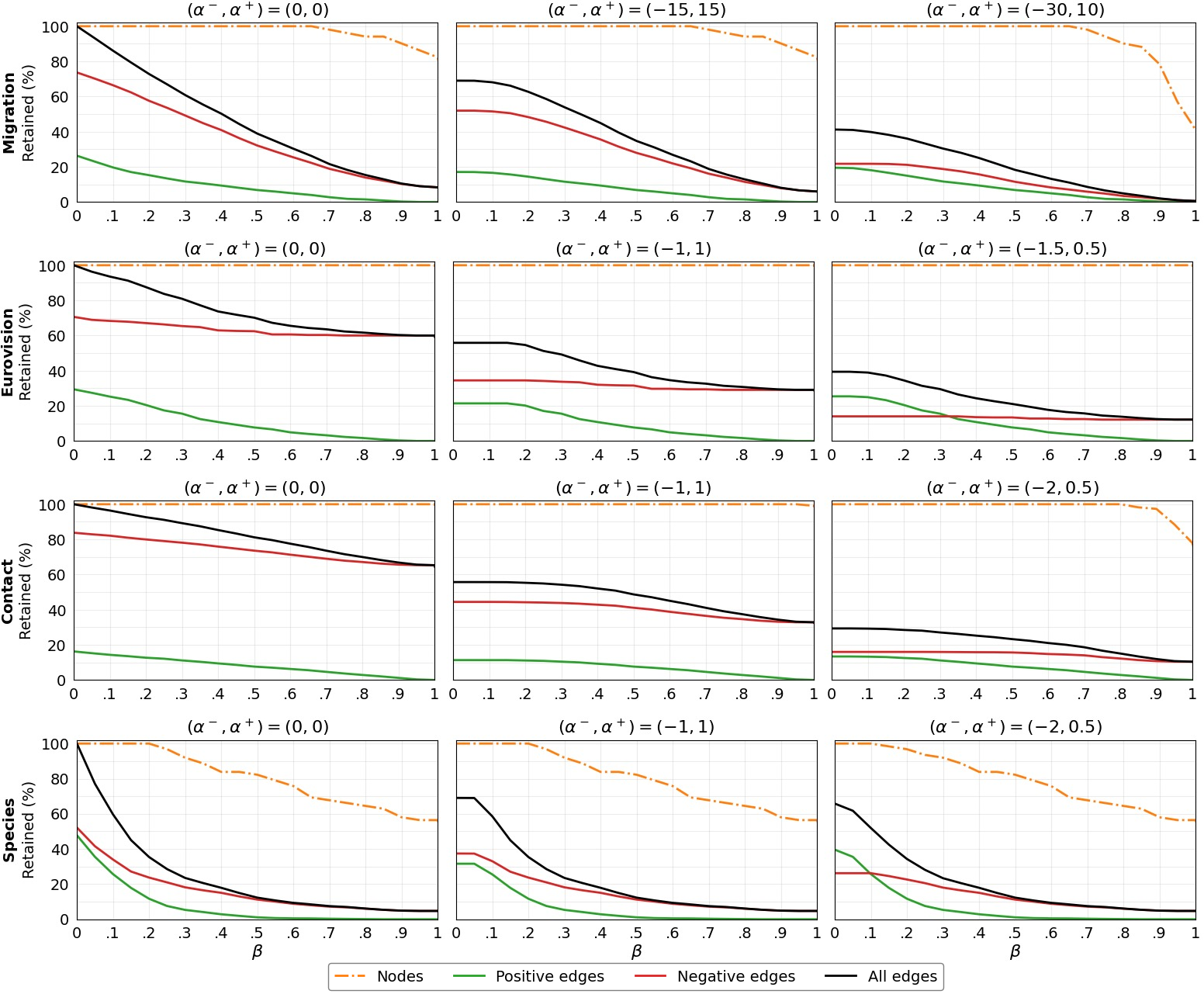}
    \caption{Effects of \textit{vigor filter} on the backbone size.}
    \label{fig:retainbeta}
\end{figure}

\textbf{Effects of vigor threshold.} In a similar fashion, we have also explored the whole continuous range of vigor threshold under different significance threshold filters in Figure \ref{fig:retainbeta}. The vigor threshold is represented with $\beta = \beta^+ = -\beta^-$ on \textit{x-axes}. Three different significance threshold settings are represented in columns. Except for \textit{Species} network and except under strict vigor thresholds, all or almost all nodes remain in the backbone. When edges are inspected, in line with the relevant conclusions derived from Figure \ref{fig:retainalpha}, increasing $\beta$ does not sufficiently eliminate the negative edges that are otherwise statistically insignificant. In very large values of $\beta$, positive links are largely eliminated. When $\beta = 1$, the edges remaining in the backbone are the zero-weighted (i.e., nonexistent) edges. Accordingly, employing very large vigor thresholds should be avoided.

\textbf{Conclusion}. Overall, we have shown that the proposed method is able to reduce an intrinsically dense network to a signed backbone of the relative size of $10\%$ to $20\%$ such that the resulting backbone contains comparable portions of negative and positive edges. Yet, the hyperparameter selection largely lies with the user and can be changed according to the purpose of analysis and the nature of the network. For instance, \textit{Species} network is constructed with similarity values as edge weights and the edge weights do not follow a skewed distribution, unlike other networks. Therefore, its behavior under the proposed filters partially deviates from the behavior of the other employed networks.

Principally, we can conclude that (i) \textit{significance filter} should almost always be used with a fairly medium threshold where fairness can be evaluated with respect to the backbone size or by inspecting the range and scale of $\alpha_{ij}$ values in the original network, (ii) very large threshold values for both filters but especially for \textit{vigor filter} should be avoided, and (iii) a sufficiently balanced, statistically meaningful, and sparse backbone can be extracted with a balanced utilization of the two proposed filters.

\subsection{Heterogeneity}
As discussed in Section \ref{sec:relwork}, an enviable backbone extraction method should respect the weight and strength heterogeneity in the original network (i.e., its multiscale, hierarchical nature). That is to say, the retained edges should not be only those originally with large weights or those connecting high-strength nodes.

\begin{figure}[h]
    \centering
    \includegraphics[width=1\textwidth]{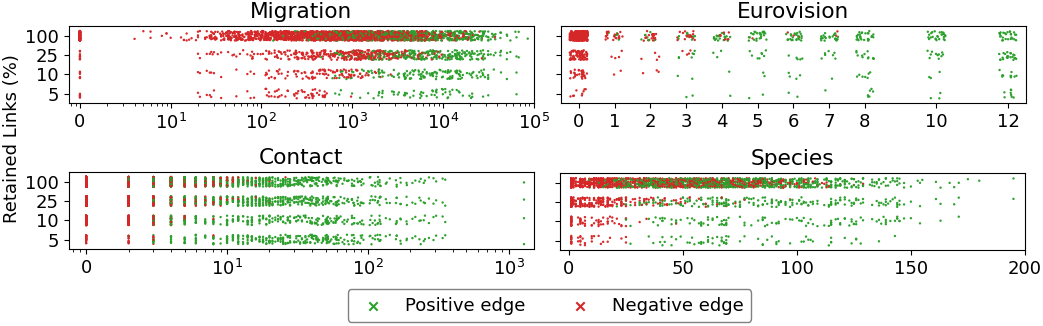}
    \caption{Original weights of the retained edges}
    \label{fig:heteroedge}
\end{figure}

\textbf{Heterogeneity of edge weights.} Figure \ref{fig:heteroedge} presents the original weight distribution\footnote{Small random noise is added on y-axis for all networks and x-axis \textit{Eurovision} network for visualization purposes.} of the retained edges in backbones of different sizes
\footnote{The backbones are extracted with the following ($\alpha^-, \alpha^+$), ($\beta^-, \beta^+$) settings for the respective backbone sizes:

\textit{Migration} $\rightarrow$ 25\%: (-33, 33), (-0.33, 0.33); 10\%: (-40, 40), (-0.57, 0.57); 5\%: (-40, 40), (-0.72, 0.72).

\textit{Eurovision} $\rightarrow$ 25\%: (-1.5, 2.25), (0, 0); 10\%: (-2, 4.5), (0, 0); 5\%: (-2.5, 5), (0, 0).

\textit{Contact} $\rightarrow$ 25\%: (-2, 2), (0, 0); 10\%: (-3, 3), (-0.66, 0.66); 5\%: (-5, 5), (-0.55, 0.55).

\textit{Species} $\rightarrow$ 25\%: (-3.5, 3.5), (0.15, 0.15); 10\%: (-5, 5), (-0.2, 0.2); 5\%: (-5.5, 5.5), (-0.33, 0.33).}
.
The figure shows that even when the extracted backbone retains only $5\%$ of all possible edges, the heterogeneity of edge weights are respected, i.e., edges with different weights are retained and some other edges with similar weights are eliminated. It also visually depicts that there is no perfect global cutoff value for determining the sign of edges as those cutoff values are established individually for each edge by the null model.

\begin{figure}[h]
    \centering
    \includegraphics[width=1\textwidth]{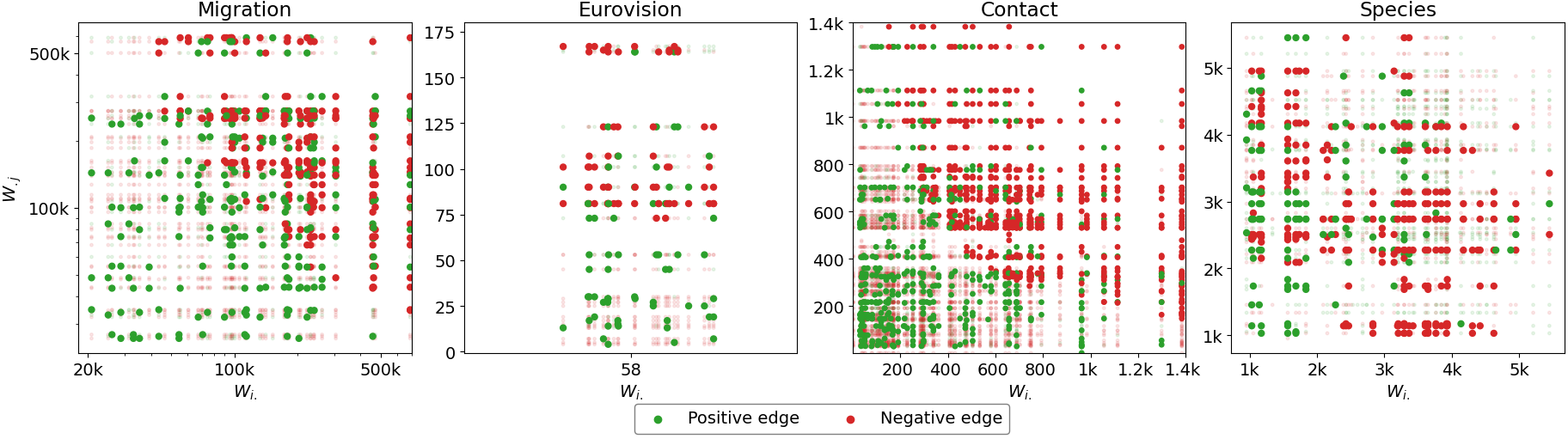}
    \caption{Original strength distribution of the retained dyads}
    \label{fig:heterostrength}
\end{figure}

\textbf{Heterogeneity of node strengths.} Figure \ref{fig:heterostrength} show the strength distribution\footnote{Small random noise is added for \textit{Eurovision} network for visualization purposes.} of the dyads for the original networks in transparent color and the extracted backbones with relative size of $\approx20\%$\footnote{The backbones in Figure \ref{fig:heteroedge} are extracted with the following ($\alpha^-, \alpha^+$), ($\beta^-, \beta^+$) settings.
\textit{Migration} $\rightarrow$ (-40, 40), (-0.25, 0.25).
\textit{Eurovision} $\rightarrow$ (-1.8, 2.4), (-0.5, 0.3).
\textit{Contact} $\rightarrow$ (-2.25, 2.25), (-0.25, 0.25).
\textit{Species} $\rightarrow$ (-3.6, 3.6), (-0.25, 0.25).}
in opaque color. Specifically, the edge between nodes $i$ and $j$ is represented with a point colored based on its sign. \textit{X-axis} denotes the (out-)strength of $i$ and \textit{y-axis} denotes the (in-)strength of $j$\footnote{For an undirected network, the order of $i$ and $j$ for an edge is established alphabetically.}. As demonstrated by the figure, the retained edges are rather evenly distributed in the plotting space and the original heterogeneity is respected. The retained edges between very low-strength nodes are usually positive since null expectations are usually low in that regime and it is difficult to conclude whether small-weighted edges in low-strength regimes are statistically significant. Likewise, the edges between very high-strength nodes are usually negative due to the upper bound on vigor implied by the large null expectation and the fixed amount of total weights. On the other hand, in the majority of strength-strength regions, the edge signs exhibit a mixed distribution.
Therefore, this is generally a desired property since the proposed method, by itself, does not allow deducing strong conclusions when the data is limited.

\textbf{Conclusion.} We can conclude that the heterogeneity given by (i) the multiscale nature of edge weights and (ii) heterogeneous nature of node strengths are respected, and (iii) the proposed filters do not draw strong conclusions when the data is limited.

\subsection{Structure of the backbones}

Generally speaking, we expect real-world networks to exhibit some extent of reciprocity \cite{DOREIAN20091, bokdd2015}, have their reciprocal edges be of the same sign in the case of signed networks, and demonstrate sufficient levels of structural balance. Moreover, most real-world networks exhibit a community structure regardless of whether they are signed or not. Therefore, an informative signed backbone should have these characteristics in general and up to a certain extent.
We analyze the extracted backbones\footnote{The backbones in Figure \ref{fig:heterostrength} are extracted with the following ($\alpha^-, \alpha^+$), ($\beta^-, \beta^+$) settings.
\textit{Migration} $\rightarrow$ (-40, 40), (-0.33, 0.33).
\textit{Eurovision} $\rightarrow$ (-1, 0.5), (-0.33, 0.1).
\textit{Contact} $\rightarrow$ (-3, 3), (-0.33, 0.33).
\textit{Species} $\rightarrow$ (-5,3), (-0.5, 0.1).}
of the employed networks in terms of reciprocity, structural balance, and community structure.

\textbf{Reciprocity.} Reciprocity in unsigned networks is generally defined as the ratio of the number of reciprocal edges to the number of all edges. In signed networks, however, the edge signs should also be considered in the analyses. Table \ref{tab:dir} presents the number of nodes, edges, nonreciprocal edges (\{$\times$, $\cdot$\}), reciprocated positive edges (\{+, +\}), reciprocated negative edges (\{-, -\}), and reciprocated edges with sign conflict (\{+, -\}) for the directed backbones extracted from directed networks. In \textit{Migration} backbone, there is considerable reciprocity with no conflicting edge signs between any node pair. In contrast, \textit{Eurovision} backbone discloses a substantial amount of conflicting edge pairs which might be of interest for further investigation into its voting dynamics (e.g., the influence of diasporas).

\begin{table}[h]
\small
\centering
\setlength{\tabcolsep}{0.5em}
\caption{Reciprocity in extracted directed backbones}
\label{tab:dir}
\begin{tabular}{l|l|l|l|l|l|l|}
\cline{2-7}
 &
  \cellcolor[HTML]{EFEFEF}\textbf{nodes} &
  \cellcolor[HTML]{EFEFEF}\textbf{edges} &
  \cellcolor[HTML]{EFEFEF}\textbf{\footnotesize{\{$\times$, $\cdot$\}}} &
  \cellcolor[HTML]{EFEFEF}\textbf{\footnotesize{\{+, +\}}} &
  \cellcolor[HTML]{EFEFEF}\textbf{\footnotesize{\{-, -\}}} &
  \cellcolor[HTML]{EFEFEF}\textbf{\footnotesize{\{+, -\}}} \\ \hline
\multicolumn{1}{|l|}{\cellcolor[HTML]{EFEFEF}\textbf{Migration}} &
  51 &
  471 &
  179 &
  188 &
  104 &
  0 \\ \hline
\multicolumn{1}{|l|}{\cellcolor[HTML]{EFEFEF}\textbf{Eurovision}} &
  26 &
  380 &
  158 &
  58 &
  80 &
  84 \\ \hline
\end{tabular}
\end{table}

For the rest of the analysis, the directed backbones are transformed into undirected backbones in the following way. For \textit{Migration} backbone, directed edges are transformed into undirected edges of the same sign with positive signs having a priority over negative signs when in conflict. For \textit{Eurovision} backbone, a positive (or negative) edge is created between two nodes when there are exactly two positive (or negative) directed edges between them (i.e., reciprocal edges of the same sign).

\textbf{Structural Balance.} As put forward by Heider \cite{heider} and formalized for signed networks by Cartwright and Harary \cite{cartwright}, an undirected triple is said to be balanced if its edges have the signs \{+, +, +\} or \{+, -, -\} and unbalanced if its edges have the signs \{+, +, -\} or \{-, -, -\}. The balanced triples can be simply described with the following expressions: \textit{friend of my friend is my friend} and \textit{enemy of my friend is my enemy}. Davis \cite{davis} defines a weaker notion and proposes that \textit{enemy of my friend is my enemy} is not necessarily required for balance and the only unbalanced triple among the four possible settings is the one with edge signs \{+, +, -\}. Accordingly, structural balance (SB) and weak structural balance (WSB) of a network can be defined as the ratio of the number of balanced triples to the number of all triples. There is also recent evidence for the existence of (weak) structural balance in social networks and voting networks \cite{Levorato2017} and spatial ecological networks \cite{sbspatial}.

Table \ref{tab:undirbackbone} shows the count of nodes, edges, and four possible triple settings; and SB and WSB measures for the undirected backbones. The backbones for \textit{Migration} and \textit{Species} shows a strong balance for all three measures. The other two backbones are also highly balanced in terms of weak structural balance which is shown to be more appropriate in real-world networks.

\begin{table}[h]
\small
\caption{Structural characteristics of extracted backbones}
\label{tab:undirbackbone}
\centering
\setlength{\tabcolsep}{0.5em} 
\begin{tabular}{l|r|r|r|r|r|r|r|r|}
\cline{2-9}
 &
  \cellcolor[HTML]{EFEFEF}\textbf{nodes} &
  \cellcolor[HTML]{EFEFEF}\textbf{edges} &
  \cellcolor[HTML]{EFEFEF}\textbf{\footnotesize{\{+, +, +\}}} &
  \cellcolor[HTML]{EFEFEF}\textbf{\footnotesize{\{+, +, -\}}} &
  \cellcolor[HTML]{EFEFEF}\textbf{\footnotesize{\{+, -, -\}}} &
  \cellcolor[HTML]{EFEFEF}\textbf{\footnotesize{\{-, -, -\}}} &
  \cellcolor[HTML]{EFEFEF}\textbf{SB} &
  \cellcolor[HTML]{EFEFEF}\textbf{WSB} \\ \hline
\multicolumn{1}{|l|}{\cellcolor[HTML]{EFEFEF}\textbf{Migration}}  & 51  & 325 & 117 & 12 & 404 & 98  & 0.83 & 0.98 \\ \hline
\multicolumn{1}{|l|}{\cellcolor[HTML]{EFEFEF}\textbf{Eurovision}} & 24  & 69  & 4   & 6  & 36  & 18  & 0.63 & 0.91 \\ \hline
\multicolumn{1}{|l|}{\cellcolor[HTML]{EFEFEF}\textbf{Contact}}    & 113 & 760 & 194 & 98 & 809 & 880 & 0.51 & 0.95 \\ \hline
\multicolumn{1}{|l|}{\cellcolor[HTML]{EFEFEF}\textbf{Species}}    & 56  & 328 & 655 & 1  & 295 & 0   & 1.00 & 1.00 \\ \hline
\end{tabular}
\end{table}

\begin{figure}[h]
\begin{subfigure}{.235\textwidth}
  \centering
  \includegraphics[width=1\linewidth]{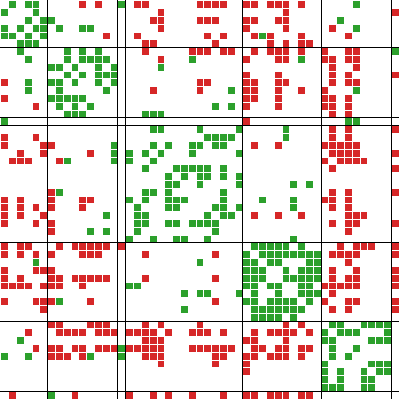}
  \caption{Migration}
  \label{fig:sfig1}
\end{subfigure}
\begin{subfigure}{.235\textwidth}
  \centering
  \includegraphics[width=1\linewidth]{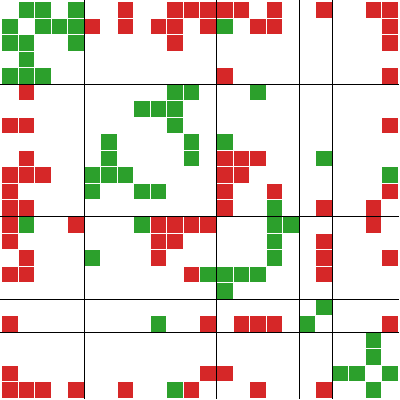}
  \caption{Eurovision}
  \label{fig:sfig2}
\end{subfigure}
\begin{subfigure}{.235\textwidth}
  \centering
  \includegraphics[width=1\linewidth]{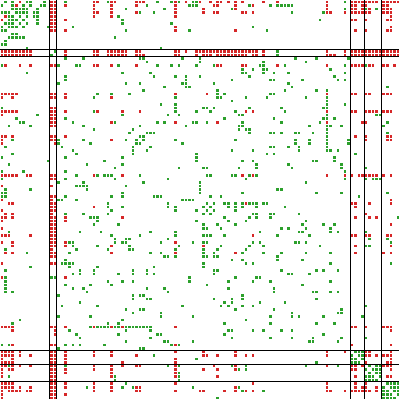}
  \caption{Contact}
  \label{fig:sfig3}
\end{subfigure}
\begin{subfigure}{.235\textwidth}
  \centering
  \includegraphics[width=1\linewidth]{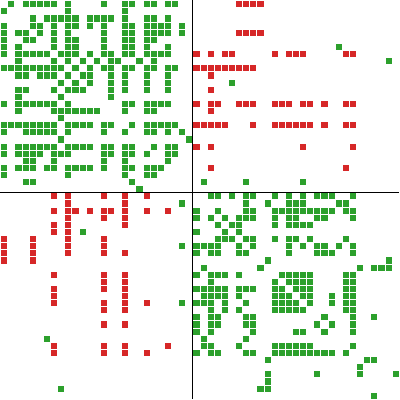}
  \caption{Species}
  \label{fig:sfig4}
\end{subfigure}

\caption{Block models of extracted backbones}
\label{fig:blocks}
\end{figure}

\textbf{Community Structure.} Based on the partitioning methods for signed networks \cite{DOREIAN20091}, the extracted backbones are partitioned into communities where the number of negative edges within communities and number of positive edges between communities are minimized via heuristics\footnote{Communities are found and visualized using \textit{signnet} \cite{signnet} package in R}. Figure \ref{fig:blocks} depicts the community structure via a visual block matrix where the order of nodes is the same for its rows and columns. Diagonal blocks are expected to cover the positive edges and non-diagonal blocks are expected to cover the negative edges. Overall, the extracted backbones manifest community structures. The backbone of \textit{Migration} has 5 clear densely connected communities and two outlier individual nodes whereas the communities in \textit{Eurovision} are not as dense. The backbone of \textit{Contact} consists of one large sparse group and several smaller and denser groups. The backbone of \textit{Species} has a very clear structure with two similar-sized communities, visually confirming the very high SB and WSB measures obtained for it.

Table \ref{tab:blocks} shows members of each group in the order of diagonal blocks where nodes are also listed in the same order as they appear in the rows and columns. In \textit{Migration} backbone, we observe that the states are grouped mostly based on their geographical proximity, which means that there are more positive links between geographically close states and more negative links between geographically distant states. This is inline with the same geographical constraints observed by several other backboning studies \cite{Serrano2009backbone, dianati2016hairball, Marcaccioli2019}. Florida, one of the two singletons here, is also shown to be the last state to connect to the rest of the network in \cite{gunter2005}. In \textit{Eurovision} backbone, for the most part, countries are grouped mainly based on geographic, cultural, linguistic, ethnic, and historical ties. In \textit{Species} backbone, the first group is composed of inshore/nearshore/estuarine species while the second one includes more Tropical/Caribbean/Reef Fish species, agreeing with the expert knowledge in marine biology. The plausibility of the found communities suggests that the proposed backbone extraction method retains the appropriate edges that preserve and reveal the important structural information available in the original network.

\begin{table}
\caption{Block memberships}
\label{tab:blocks}
\setlength{\tabcolsep}{1pt}
\footnotesize
\centering
\begin{tabular}{|l|l|l|} 
\hline
\rowcolor[rgb]{0.937,0.937,0.937} \textbf{Migration}       & \textbf{Eurovision}                 & \textbf{Species}                                                                                                          \\ 
\hline
\begin{tabular}[c]{@{}l@{}}Maryland, D. of Columbia., \\S. Carolina, W. Virginia,\\Virginia, N. Carolina \end{tabular}                                                                          & \begin{tabular}[c]{@{}l@{}}Bosnia and Herzegovina, \\Croatia, Turkey, \\Slovenia, Austria \end{tabular}                            & \multirow{4}{*}{\begin{tabular}[c]{@{}l@{}}Florida Gar, Gray Snapper, Gulf Flounder,\\ Ladyfish, Mangrove Species Rivulus, \\ Oyster Toadfish, Sheepshead, Red Drum, \\ Florida Pompano, American Eel, Brown Shrimp,\\ Bucktooth Parrotfish, Pink Shrimp, Spanish Sardine, \\ Sheepshead Minnow, Bonefish, Permit, Striped Mullet,\\ Snook, Black Tip Shark, Spotted Sea Trout, Tarpon, \\ Yellowfin Mojarra, Pinfish, Crevalle Jack, \\ Goliath Grouper, Lemon Shark \end{tabular}}                                                                                                  \\ 
\cline{1-2}
\begin{tabular}[c]{@{}l@{}}Georgia, Kentucky, Louisiana, \\ Mississippi, Tennessee, Ohio, \\ Alabama, Indiana, Arkansas \end{tabular}                                                           & \multirow{2}{*}{\begin{tabular}[c]{@{}l@{}}Belgium, France, Germany, \\ Portugal, Romania, Poland, \\ Spain, Sweden \end{tabular}} &                               

\\ 
\cline{1-1}
Delaware
&
&

\\ 
\cline{1-2}
\begin{tabular}[c]{@{}l@{}}New Mexico, Wisconsin,\\Missouri, Oklahoma, Texas,\\Nebraska, S. Dakota, Wyoming, \\Kansas, Colorado, Minnesota,\\Illinois, Iowa,\\Michigan, N. Dakota \end{tabular} & \begin{tabular}[c]{@{}l@{}}Ireland, Iceland, Netherlands, \\ Norway, Malta \end{tabular}                                           &
\\ 
\hline
\begin{tabular}[c]{@{}l@{}}California, Oregon, \\Hawaii, Nevada, \\Washington, Arizona, \\Montana, Idaho, \\ Utah, Alaska \end{tabular}                                                         & Cyprus, Greece                                                                                                                     & \multirow{3}{*}{\begin{tabular}[c]{@{}l@{}}Bluehead Wrasse, Carribean Spiny Lobster, \\French Angelfish, French Grunt, Hogfish, Jolthead Porgy, \\Long-Spined Sea Urchin, Queen Conch, Queen Triggerfish,\\King Mackerel, Atlantic Sailfish,~Black Grouper, \\Dusky Squirrelfish, Longsnout Butterflyfish, Sergeant Major,\\Spanish Spotted Lobster, Barred Hamlet, \\Carribean Reef Squid, Green Moray, Reef Croker, \\Florida Stone Crab, Black Margate, Cero Mackerel, \\Tripletail Peacock Flounder, Stoplight Parrotfish, \\ Bar Jack, Cobia, Yellowtail Snapper \end{tabular}}  \\ 
\cline{1-2}
\begin{tabular}[c]{@{}l@{}}Massachusetts, New York, \\ Connecticut, New Jersey, \\ Pennsylvania, \\New Hampshire, Maine, \\Rhode Island, Vermont \end{tabular}                                  & \multirow{2}{*}{\begin{tabular}[c]{@{}l@{}}Estonia, Israel, \\ Russia, Ukraine \end{tabular}}                                      &

\\ 
\cline{1-1}
\begin{tabular}[c]{@{}l@{}}Florida \\\\\end{tabular}
&
&

\\
\hline
\end{tabular}
\end{table}

\textbf{Conclusion.} We showed that the extracted backbones usually (i) reveal interesting or expected reciprocal structures, (ii) are structurally balanced, and (iii) exhibit clear and intuitive community structure in agreement with the existing knowledge.

\section{Conclusion and Future Work}

In this study, we put forward an initial discussion on intrinsically dense networks and provide \textit{significance filter} and \textit{vigor filter} for extracting signed backbones of such networks. Empirical evaluations that utilize the proposed filters on a variety of real-world networks show that sparse backbones can be obtained while maintaining comparable numbers of positive and negative links and respecting the original weight and strength heterogeneity. In general, the obtained backbones exhibits characteristics associated with signed networks such as reciprocity, structural balance, and community structure. The extraction method, by design, does not prematurely arrive at conclusions regarding the existence of signed links between nodes when the data is limited. On the other hand, choosing appropriate hyperparameter values lies with the user and should be guided by empirical analysis and recommendations provided in this work as well as the purpose and the nature of the specific problem.

Further studies can improve this work in two major ways. As we presented many examples throughout the manuscript, intrinsically dense networks exist in many different domains. First, a stream of its applications in different fields and utilization of the resulting backbones in different tasks would provide important feedback regarding its usefulness, weaknesses, and strengths. Second, hypergeometric distribution characterizes a discrete process and such discrete distributions are being utilized in the state-of-the-art backbone extraction methods including ours. As we discussed earlier, null models developed based on it can have certain undesirable properties for certain networks. Therefore, a natural avenue is to develop new null models that are more appropriate for the case of link weights that are continuous or that can be equivalently represented in different units.

\newpage
\section*{Appendix A. The Software Packages}\label{app:package}

The code for the extraction methods proposed in this study is published as an MIT-licensed Python package \textit{signed\_backbones}\footnote{The Python package is available on the Python Package Index at https://pypi.org.} and an MIT-licensed R package \textit{signed.backbones}\footnote{The R package is currently available at https://github.com/furkangursoy/signed.backbones. We anticipate that it will be available on CRAN before the possible publication of this manuscript.}.

In Python, the package can be installed using the package manager \textit{pip} as follows.

\begin{lstlisting}[breaklines, basicstyle=\footnotesize]
pip install signed_backbones
\end{lstlisting}

In R, (at this time) the package can be installed using the package \textit{devtools} as follows.

\begin{lstlisting}[breaklines, basicstyle=\footnotesize]
devtools::install_github("furkangursoy/signed.backbones")
\end{lstlisting}

In both packages, the filters are provided through a function named \textit{extract()}. Its parameters are the same for the packages in both languages and are summarized below. Language-specific details are documented within the respective packages.
\begin{itemize}[itemsep=0pt]
    \item \textbf{\textit{edgelist}}: First two columns contain node pairs, and the third column contains the edge weights. If \textit{directed = True}, columns should be in this order: source node, target node, edge weight.
    
    \item \textbf{\textit{directed}}: \textit{True} or \textit{False} for indicating whether the input network is directed.

    \item \textbf{\textit{significance\_threshold}}: Threshold for the significance filter. If filtering is based directly on $\alpha$ values: a single nonnegative value, e.g., 1.23; or a couple of nonpositive and nonnegative values, e.g., (-1.23, 4.56). If filtering is based on ranking: a single percentage value in the format such as '10pc'; or a couple of percentage values in the format such as ('5pc', '5pc').
    
    \item \textbf{\textit{vigor\_threshold}}: Threshold for the vigor filter. A single nonnegative value in the range [0, 1], e.g., 0.33; or a couple of nonpositive and nonnegative values in the ranges [-1, 0] and [0, 1], e.g., (-0.5, 0.3).
    
    \item \textbf{\textit{return\_weights}}: Whether the returned backbone should contain the signed link weights that show the intensity of the link sentiment.
    
    \item \textbf{\textit{return\_significance}}: Whether the returned backbone should contain the link significance values that are benchmarked against the \textit{significance\_threshold}.
    
    \item \textbf{\textit{max\_iteration}}: Maximum number of iterations to be used in the IPFP.
    
    \item \textbf{\textit{precision}}: A small epsilon value for numerical precision issues. It can be left as default.
\end{itemize}

The \textit{extract()} function returns a tabular data where each row represents a link in the extracted backbone. First two columns contain node pairs, and the third column contains the edge sign. If \textit{directed = True}, columns are in this order: source node, target node, edge sign. If \textit{return\_weights = True}, signed edge weights ($\beta_{ij}$) are returned instead of edge sign. If \textit{return\_significance = True}, a fourth column containing significance values ($\alpha_{ij}$) are returned.

The following piece of code in Python extracts the signed backbone of an undirected network using significance filter threshold $-\alpha^- = \alpha^+ = 2.576$ and vigor filter threshold $(\beta^-, \beta^+) = (-0.3, 0.2)$. The variable \textit{sbb} contains the signed edges of the extracted backbone.

\begin{lstlisting}[breaklines, basicstyle=\footnotesize]
import signed_backbones as sb
import pandas as pd
net = pd.read_csv('edgelist.txt', ...)
sbb = sb.extract(net, directed = False, significance_threshold = 2.576, vigor_threshold = (-0.3, 0.2))
\end{lstlisting}

The following piece of code in R extracts the signed backbone of a directed network using significance filter threshold $(\alpha^- = \alpha^+) = (\text{'10pc', '10pc'})$ and vigor filter threshold $-\beta^- = \beta^+ = 0.1$. The variable \textit{sbb} contains the signed and weighted edges of the extracted backbone.

\begin{lstlisting}[breaklines, basicstyle=\footnotesize]
net <- read.csv(edgelist.txt', ...)
sbb <- signed.backbones::extract(net, directed = TRUE, significance_threshold = c('10pc', '10pc'), vigor_threshold = 0.1, return_weights = TRUE)
\end{lstlisting}

\newpage

\section*{Appendix B. Hypergeometric Distribution Intervals}\label{app:interval}

\begin{figure}[h]

\begin{subfigure}{.48\textwidth}
  \centering
  \includegraphics[width=1\linewidth]{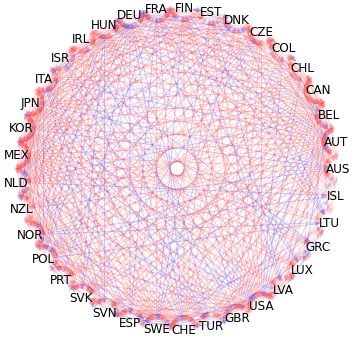}
  \caption{When unit = 10M USD}
  \label{fig:sfig10m}
\end{subfigure}
\begin{subfigure}{.48\textwidth}
  \centering
  \includegraphics[width=1\linewidth]{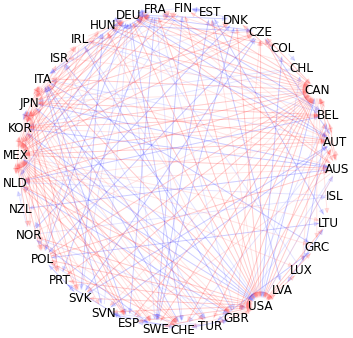}
  \caption{When unit = 100M USD}
  \label{fig:sfig100m}
\end{subfigure}

\caption{Hypergeometric backbones of the OECD trade network based on traditional hypothesis testing}
\label{fig:purehg}
\end{figure}

Testing statistical significance of edge weights with hypergeometric distribution in a traditional way is not appropriate when edge weights are not counts of discrete events but continuous values with possibly arbitrary units. Here, we employ the trade network between OECD countries in 2019\footnote{The data is obtained from \textit{OECD.Stat} website at https://stats.oecd.org/Index.aspx?DataSetCode=BTDIXE\_I4.} where edge weights represent the amount of exports in monetary amounts. We show that when edge weights are represented in units of 1 million USD and 10 million USD, the confidence intervals obtained from hypergeometric distribution change. Table \ref{tab:ci} presents actual edge weights and confidence intervals with \%99.999 confidence for the two cases. Randomly selected 5 edges are included and all values are shown in 1000 USDs. Figure \ref{fig:purehg} visualizes the networks extracted based on the same confidence value. The density of extracted backbones are 0.65 and 0.28 respectively in Figure \ref{fig:sfig10m} and \ref{fig:sfig100m}. As shown by these results, when edge weight units do not correspond to discrete events, traditional hypothesis testing with such discrete distributions does not generalize well to the continuous edge weights.

\begin{table}[h!]
\small
\centering
\setlength{\tabcolsep}{0.5em}
\caption{Confidence intervals in the OECD trade network}
\label{tab:ci}
\begin{tabular}{llr|r|r|r|r|}
\cline{4-7}
\textbf{}                             & \textbf{}                            & \multicolumn{1}{l|}{\textbf{}} & \multicolumn{2}{c|}{\textbf{unit = 10M USD}} & \multicolumn{2}{c|}{\textbf{unit = 100M USD}} \\ \hline
\multicolumn{1}{|l|}{\textbf{Source}} & \multicolumn{1}{l|}{\textbf{Target}} & \textbf{Observed}              & \textbf{Lower Bound}  & \textbf{Upper Bound} & \textbf{Lower Bound}  & \textbf{Upper Bound}  \\ \hline
\multicolumn{1}{|l|}{ITA}             & \multicolumn{1}{l|}{BEL}             & 15,742,248.79                  & 14,900,000.0          & 18,330,000.0         & 11,400,000.0          & 22,300,000.0          \\ \hline
\multicolumn{1}{|l|}{ISL}             & \multicolumn{1}{l|}{BEL}             & 77,439.63                      & 30,000.0              & 390,000.0            & 0.0                   & 1,000,000.0           \\ \hline
\multicolumn{1}{|l|}{POL}             & \multicolumn{1}{l|}{AUT}             & 5,376,834.71                   & 3,410,000.0           & 5,190,000.0          & 1,800,000.0           & 7,400,000.0           \\ \hline
\multicolumn{1}{|l|}{CZE}             & \multicolumn{1}{l|}{CAN}             & 337,201.49                     & 7,560,000.0           & 10,090,000.0         & 5,100,000.0           & 13,000,000.0          \\ \hline
\multicolumn{1}{|l|}{NLD}             & \multicolumn{1}{l|}{AUT}             & 7,013,139.08                   & 8,360,000.0           & 10,990,000.0         & 5,800,000.0           & 14,000,000.0          \\ \hline
\end{tabular}
\end{table}

\newpage
\bibliographystyle{unsrt}
\bibliography{mybibfile}

\end{document}